\documentclass[twocolumn,prl,showpacs,superscriptaddress,showpacs]{revtex4-1}
\usepackage{amssymb,amsmath}
\usepackage{graphicx}
\usepackage{bm}
\usepackage{dcolumn}
\usepackage{float}
\usepackage{url}
\usepackage{color}
\usepackage{subfigure}
\usepackage{siunitx}
\usepackage{txfontsb}
\usepackage[OT1]{fontenc}
\usepackage[driverfallback=dvipdfm]{hyperref}
\usepackage{soul,ulem}
\hypersetup{colorlinks=true,breaklinks,urlcolor=blue,linkcolor=blue,citecolor=blue}

\begin{document}

\title{Observation of Kibble-Zurek Behavior across Topological Transitions of a Chern Band in Ultracold Atoms}
\date{\today}

\author{Huan Yuan}
\author{Chang-Rui Yi}
\author{Jia-Yu Guo}
\author{Xiang-Can Cheng}
\author{Rui-Heng Jiao}
\author{Jinyi Zhang}
\author{Shuai Chen}
\author{Jian-Wei Pan}
\affiliation{Hefei National Research Center for Physical Sciences at the Microscale and School of Physical Sciences, University of Science and Technology of China, Hefei 230026, China}
\affiliation{Shanghai Research Center for Quantum Science and CAS Center for Excellence in Quantum Information and Quantum Physics, University of Science and Technology of China, Shanghai 201315, China}
\affiliation{Hefei National Laboratory, University of Science and Technology of China, Hefei 230088, China}

\begin{abstract}
The Kibble-Zurek (KZ) mechanism renders a theoretical framework for elucidating the formation of topological defects across continuous phase transitions.
Nevertheless, it is not immediately clear whether the KZ mechanism applies to topological phase transitions. 
The direct experimental study for such a topic is hindered by quenching a certain parameter over orders of magnitude in topological materials.
Instead, we investigate the KZ behavior across topological transitions of a Chern band in two-dimensional (2D) optical Raman lattices with quantum gases. Defined as the defects, excitation density is reconstructed via measuring the spin wave functions, with which the power-law scaling of total excitation density is extracted and such scaling could be interpreted within the KZ framework. 
Our work has heralded the commencement of experimentally exploring the KZ mechanism of the topological phase transitions.

\end{abstract}

\maketitle
The famous Kibble-Zurek (KZ) mechanism \cite{RN346,RN347}
portrays the formation of topological defects when a system is driven across a continuous phase transition at a finite rate and then the density of defects possesses universal power-law scalings.
Such mechanism has been experimentally investigated for both classical and quantum phase transitions in a wide variety of systems, such as cosmic microwave background \cite{PhysRevLett.100.021301}, superconductor \cite{PhysRevLett.84.4966,PhysRevLett.89.080603}, and ultracold atoms \cite{doi:10.1126/science.1258676,doi:10.1126/science.aaf9657,RN391,nphys2734}.
On the other hand, in the last two decades, great interest in topological physics in solid-state materials, photonic crystals, and ultracold atoms brought about cutting-edge studies of new topological matters \cite{RevModPhys.82.3045,RevModPhys.83.1057}.
Topological phases are identified by global topological invariant, and thus, their description falls outside the Landau paradigm with the concept of local order parameters and spontaneous symmetry breaking.

When studying dynamics for topological phases, people could ask whether the KZ mechanism works in topological systems.
Currently, to address such questions, there exist numerical studies based on different models, which could be broadly classified into two categories. 
The first category investigates the relationship between the bulk excitation density and the quench rate of a certain parameter, meanwhile discovers that the excitation density versus the quench rate exhibits a KZ power-law scaling \cite{PhysRevLett.102.135702,PhysRevB.97.235144,PhysRevA.103.013314,PhysRevB.100.125110}.
The second category starts with the Berry curvature in momentum space and further defines the correlation length of the topological system to determine whether the correlation length versus quench rate satisfies the KZ mechanism \cite{PhysRevLett.125.216601,PhysRevB.106.134203,PhysRevB.110.165130,Chen_2019,PhysRevB.84.241106,PhysRevLett.114.056801}.
To date, the direct experimental measurement of KZ behavior across topological phase transitions in topological materials is impeded by the following two ingredients: i) precisely tuning a certain parameter at a finite quench rate across the transition point with a wide range (at least one order of magnitude). 
ii) proper measurement method for extracting observables such as the topological defects. 
Instead, ultracold atoms can provide a highly controllable platform for studying the dynamics of a topological system. 
Especially, the advent of the optical Raman lattices with spin-orbit coupled quantum gases \cite{PhysRevLett.112.086401,PhysRevLett.121.150401,doi:10.1126/science.aaf6689} renders a versatile platform for exploring the KZ behavior in topological systems of ultracold atoms.
In this system, for two dimensional (2D) spin-orbit coupled quantum gas, sudden quench dynamics have been performed to characterize the band topology \cite{PhysRevLett.121.250403,ZHANG20181385,PhysRevLett.118.185701,PhysRevLett.123.190603}.
In addition, for 1D spin-orbit coupled BEC, the homogeneous and inhomogeneous KZ mechanism \cite{PhysRevLett.125.260603} also has been simultaneously manifested.

\begin{figure*}
  \centering
  \includegraphics[width=0.9\linewidth]{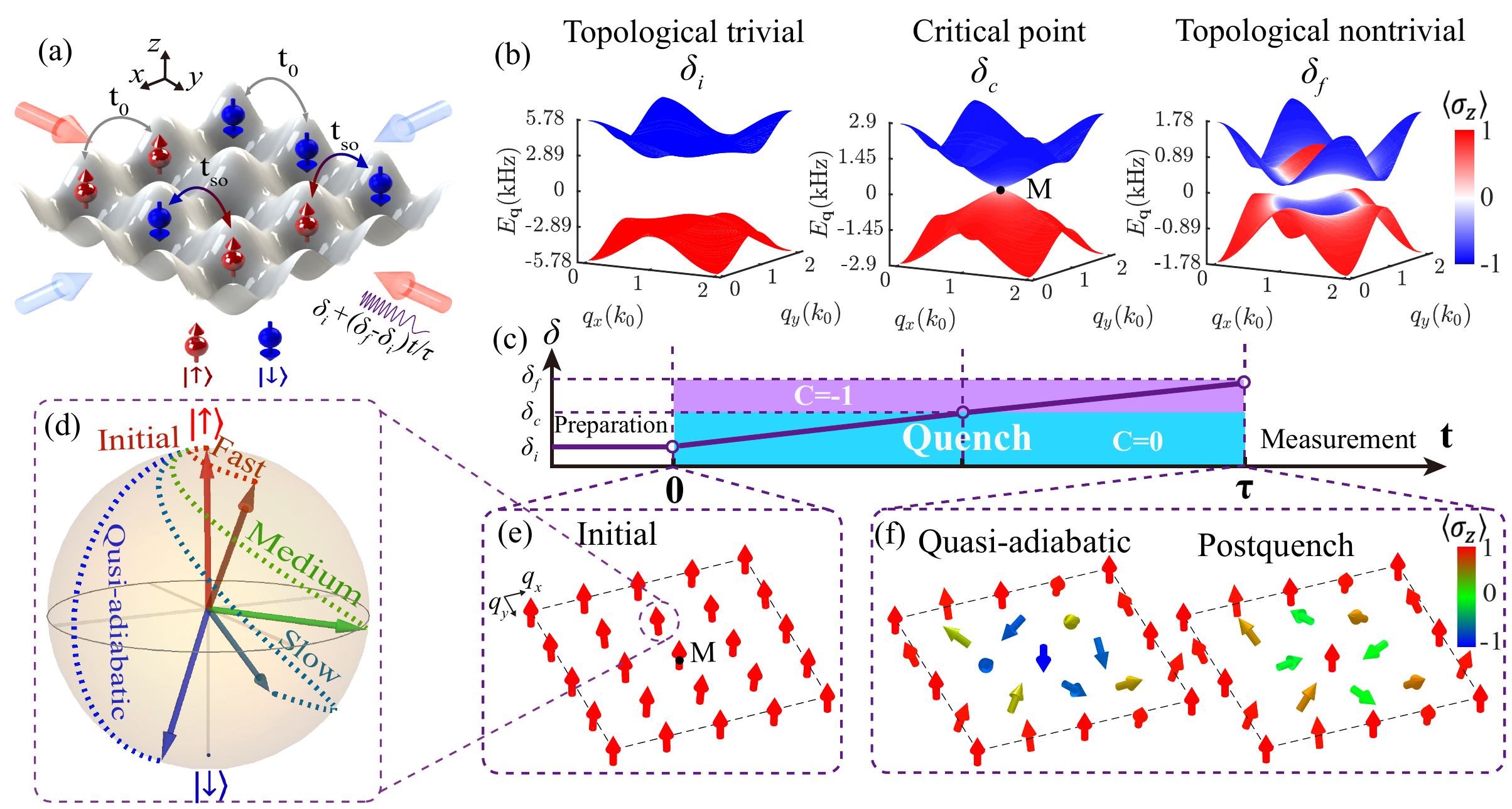}\\
  \caption{Quench across topological transitions of a Chern band in 2D optical Raman lattices.
  (a) Sketch map of Raman lattices.
  (b) Band structures in the topologically nontrivial (trivial) region with the $\langle\sigma_{z}\rangle$ presented.
  (c) Schematics of the experimental quench protocol.
  (d) The Bloch sphere displays the evolution of the Bloch vectors at quasimomentum $\mathbf{q}=(\pi, 0.7\pi)$ for different quench time.
  (e) The cartoon distribution of initial state Bloch vectors $\langle\bm{\sigma}\rangle$ in quasimomentum space.
  (f) The cartoon distribution of postquench and quasiadiabatic Bloch vectors $\langle\bm{\sigma}\rangle$ in quasimomentum space and the color of arrows denote $\langle\sigma_{z}\rangle$.}\label{Fig1}
\end{figure*}

Here, we experimentally observe a power-law scaling of total excitation density across topological transitions of a Chern band in 2D optical Raman lattices.
Based on the Bloch state tomography \cite{PhysRevResearch.5.L032016}, the spin wave functions during quench dynamics are directly observed via the measurement of all components of the Bloch vectors.
Further, the momentum distribution of excitation density, defined by the spin wave functions, is reconstructed.
Thus, the power law scaling of the total excitation density versus quench rate is obtained, which demonstrates the KZ behavior in the topological transition of the Chern band. The approach is broadly applicable and can be realized on various physical platforms beyond our system \cite{doi:10.1126/sciadv.aao4748,PhysRevLett.132.183802}.

\begin{figure}
  \centering
  \includegraphics[width=0.9\linewidth]{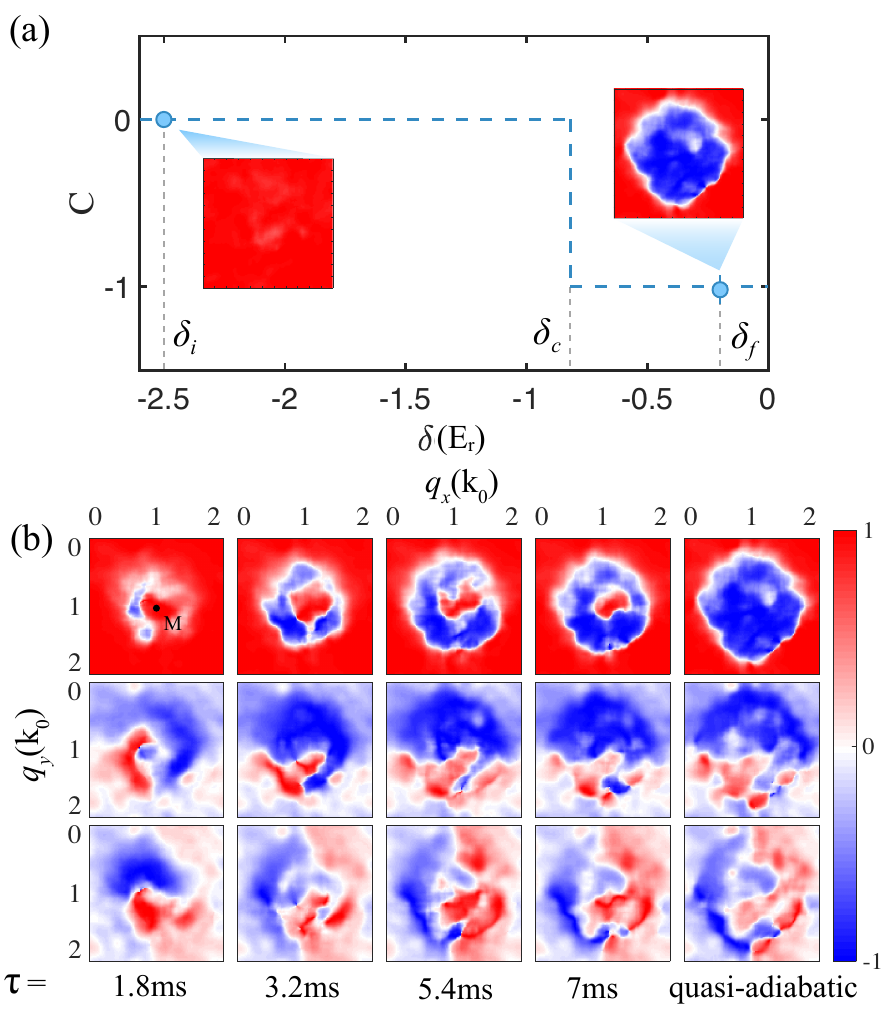}\\
  \caption{Normalized distribution of expectation values of Pauli matrices $\langle\sigma_{x,y,z}\rangle$ after the quench in the first Brillouin zone (FBZ) for different quench time $\tau$ and quasiadiabatic process.
  (a) The experimentally measured initial Chern number $C=0$ for $\delta_i=-2.5E_{\rm{r}}$ and Chern number $C=-1.02\pm0.09$ for $\delta_f=-0.2E_{\rm{r}}$.
  The corresponding distribution of $\langle\sigma_{z}(\mathbf{q})\rangle$ is presented in the inset.
  (b) Normalized distribution of expectation values of Pauli matrices $\langle\sigma_{x,y,z}\rangle$ after the quench in the FBZ for different quench time $\tau$ and quasiadiabatic process.
  The first row represents $\langle\sigma_{z}\rangle$, the second row represents $\langle\sigma_{x}\rangle$, and the third row represents $\langle\sigma_{y}\rangle$. The data are normalized by $\sqrt{\langle \sigma_{x}(\mathbf{q}) \rangle^2+\langle \sigma_{y}(\mathbf{q}) \rangle^2+\langle \sigma_{z}(\mathbf{q}) \rangle^2}=1$ with the smoothing and the smoothing is achieved by averaging over a $9\times9$ window, a region that is very small compared to the entire first Brillouin zone of size $109\times109$, accounting for only $0.68\%$ of its area.
  Parameters: $(V_{\text{0}},\Omega_{\text{0}},\delta_i,\delta_f)=(4.0,1.0,-2.5,-0.2)E_{\rm{r}}$.
  }\label{Fig2}
\end{figure}

Our experiments start with the quantum anomalous Hall (QAH) model, which is the two-band tight-binding approximation of the 2D Raman lattices for the ultracold $^{87}\text{Rb}$ atoms \cite{PhysRevLett.121.150401,supM} [Fig.\ref{Fig1} (a)].
The Raman lattices are constructed with two counterpropagating laser beams.
The two beams couple two magnetic sublevels $\left| F=1,m_F=-1 \right \rangle$(spin up $\left| \uparrow \right \rangle$) and $\left| F=1,m_F=1 \right \rangle$ (spin down $\left| \downarrow \right \rangle$) with Raman coupling processes \cite{PhysRevLett.112.086401,PhysRevLett.121.150401,doi:10.1126/science.aaf6689}.
The Hamiltonian of the QAH model is
\begin{align}
  \begin{split}
  \mathcal{H}(\mathbf{q})&=2t_{\text{so}}\sin{q_y} \sigma_x+2t_{\text{so}}\sin{q_x} \sigma_y\\
  &+\left[\delta/2-2t_\text{0}\left( \cos{q_x} + \cos{q_y} \right)\right] \sigma_z \label{eq1}
  \end{split}
\end{align}
with quasimomentum $\mathbf{q}=\left(q_x,q_y\right)$, Pauli matrices $\bm{\sigma}=(\sigma_x,\sigma_y,\sigma_z)$, $\delta$ being two-photon detuning,
and $t_{\text{0}}$ ($t_{\text{so}}$) being the spin-conserved (spin-flipped) hopping coefficient; we set the lattice constant as the length unit.
The spin-conserved (spin-flipped) hopping coefficient $t_{\text{0}}$ ($t_{\text{so}}$) is set by the lattice depth $V_{\text{0}}$ (the Raman strength $\Omega_{\text{0}}$)~\cite{supM}.
The Hamiltonian possesses topological nontrivial (trivial) band with $0<|\delta|<8t_{\text{0}}$ ($|\delta|>8t_{\text{0}}$), which implies the topological transition of a Chern band occurs when the detuning is tuned from the initial detuning $|\delta_i|>8t_{\text{0}}$ to the final detuning $0<|\delta_f|<8t_{\text{0}}$.
In this process, the gap of the bands closes at the critical point $\delta=\delta_c$, while the gap opens at the rest of the detuning, as depicted in the upper of Fig.\ref{Fig1}(b).

Following the spirit of the KZ mechanism, we perform quench protocol across topological transitions of the Chern band by applying a linear modulation of the detuning as $\delta (t)= \delta_i + (\delta_f - \delta_i)t/\tau$.
The quench protocol is depicted in Fig.\ref{Fig1}(c).
The steps are as follows:
(1) the atoms are populated to each momentum point of the lowest band in the topologically trivial band with detuning $\delta_i = -2.5E_{\rm{r}}$ at $t=0$, in which the states are fully polarized ($\left| \uparrow \right \rangle$ state)[Fig.~\ref{Fig1}(e)].
The lattice depth $V_{\text{0}}=4E_{\rm{r}}(3E_{\rm{r}})$ and the Raman coupling strength $\Omega_{\text{0}}=1E_{\rm{r}}$ are fixed during the quench.
(2) We linearly ramp the detuning with quench time $\tau$ to cross the transition point at $\delta_c=-0.82E_{\rm{r}}$ and then the system reaches the topologically nontrivial band with $\delta_f = -0.2E_{\rm{r}}$ at $t=\tau$.
(3) To obtain the excitation behavior during the dynamical process, we measure the momentum distributions of all components of the Bloch vectors $\langle \sigma_{x,y,z}(\mathbf{q})\rangle$ at $t=\tau$ via the Bloch state tomography\cite{PhysRevResearch.5.L032016,supM}.
Concretely, $\langle \sigma_z(\mathbf{q})\rangle$ is directly observed from the spin-resolved time-of-flight (TOF) imaging, while $\langle \sigma_{x,y}(\mathbf{q})\rangle$ are measured by applying the momentum transferred $\pi/2$ Raman pulses to rotate the spin basis before TOF imaging. Compared to previous studies \cite{PhysRevLett.121.250403,PhysRevResearch.5.L032016}, we improved the stability of atom number and temperature to 6\% and $7\mathrm{nK}$, respectively, while maintaining $2\times10^{5}$  atoms at 100 $\mathrm{nK}$. We further developed dynamic Bloch-state tomography by stabilizing the relative phase between the Raman pulse and the lattice via precise control of the radio-frequency signals driving the lattice lasers. This technical advance enables reliable measurement of the Bloch vector during quench dynamics and extraction of power-law scaling.

When detuning $\delta$ is linear ramp with the quench time $\tau$ from the trivial band to the topological band, the Bloch vectors $\langle \bm{\sigma}(\mathbf{q}) \rangle$ near the band closure $M$ point $[\mathbf{q}=(\pi, \pi)]$ are evolved nonadiabatically in the Bloch sphere [Fig.~\ref{Fig1}(d)]. 
The dashed box in Fig.\ref{Fig1}(f) displays the cartoon distributions of the quenched Bloch vectors $\langle \bm{\sigma}(\mathbf{q})\rangle$ and quasiadiabatic $\langle \bm{\sigma}(\mathbf{q})\rangle$, in which a deviation around the band closure $M$ point is clearly observed.
For the quasiadiabatic process, the Bloch vectors evolve adiabatically, and thus most of the atoms remain in the lower band without excitations.
Around the $M$ point, the Bloch vectors possess the clear spin flip.  
For the finite quench rate, the Bloch vectors evolve nonadiabatically such that partial atoms are excited to the higher bands. 
Around the $M$ point, the Bloch vectors exhibit an incomplete spin flip. 
Such deviation from the quasiadiabatic states reflects the excitation behavior.

To quantify the excitations during the dynamical process, the excitation density $\rho_e(\mathbf{q},\tau)$ at quasimomentum point $\mathbf{q}$ with the quench time $\tau$ is defined as $\rho_e(\mathbf{q},\tau)=1-|\langle \Psi(\mathbf{q},\tau) | \Psi_{\rm adi}(\mathbf{q}) \rangle|^2$\cite{PhysRevLett.102.135702,PhysRevB.97.235144,PhysRevA.103.013314,PhysRevB.100.125110}, where $\Psi(\mathbf{q},\tau)$ and $\Psi_{\rm adi}(\mathbf{q})$ are the quenched and quasiadiabatic spin wave functions, respectively.
The excitation density $\rho_e(\mathbf{q},\tau)$ measures the probability density that the atoms at the qausimomentum $\mathbf{q}$ are excited away from the quasiequilibrium state (quasiadiabatic processes) with quench time $\tau$.
In the scenario of a topological band, it requires taking into account the excitation density for all quasimomentum points in the first Brillouin zone (FBZ) such that it can depict the quantification of excitations relative to the topological ground state.
The total excitation density $n_{\rm exc}(\tau)$ is thus estimated by the integral of the excitation density over the full FBZ, namely \cite{PhysRevLett.102.135702,PhysRevB.97.235144,PhysRevA.103.013314,PhysRevB.100.125110}
\begin{align}
n_{\rm exc}(\tau) \sim \int_{\text{FBZ}} \rho_e(\mathbf{q},\tau) \text{d} \mathbf{q}, \label{eq2}
\end{align}
Therefore, it is natural to use the total excitation density $n_{\rm exc}(\tau)$ to characterize the dynamical behavior across the topological transition.

\begin{figure}
  \includegraphics[width=1\linewidth]{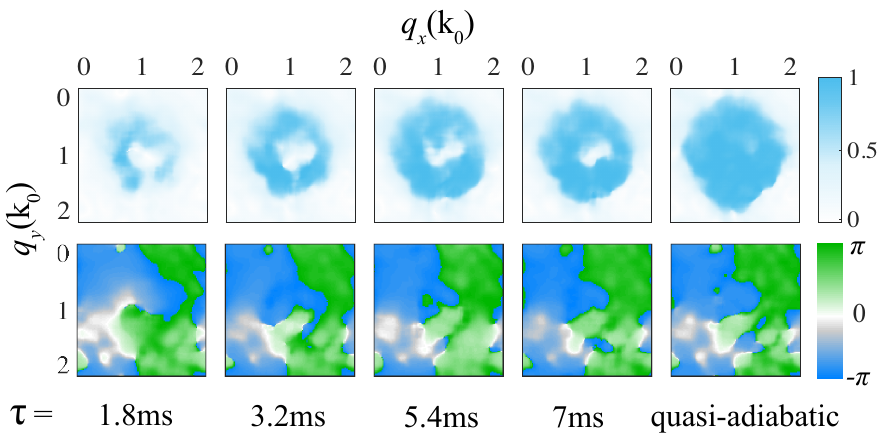}
  \caption{Reconstruction of the spin wave function in the FBZ:
  The first row represents the amplitude of the wave function $u_\downarrow$ for different quench time $\tau$ and quasiadiabatic processes.
  The second row represents the relative phase $\phi(\mathbf{q})$ between $u_\uparrow$ and $u_\downarrow$.
  Parameters: $(V_{\text{0}},\Omega_{\text{0}},\delta_i,\delta_f)=(4.0,1.0,-2.5,-0.2)E_{\rm{r}}$.
  }\label{Fig3}
\end{figure}

In terms of Eq.(\ref{eq2}), the spin wave functions are essential to be reconstructed, for which the Bloch vectors $\langle \sigma(\mathbf{q}) \rangle$ are first experimentally measured.
In Fig.\ref{Fig2}(a), According to the formula $C=-\frac{1}{4\pi}\int_{\text{FBZ}}\langle\bm{\sigma} (\mathbf{q}) \rangle \cdot (\partial_{q_x}\langle \bm{\sigma}(\mathbf{q}) \rangle \times \partial_{q_y} \langle \bm{\sigma}(\mathbf{q}) \rangle)\rm{d}\mathbf{q}$, we present the experimentally measured Chern number for initial and finial detuning.
The jump of the Chern number from $0$ to $-1$ provides experimental evidence for the topological transition of the Chern band when $\delta$ is tuned from $\delta_i$ to $\delta_f$.
Figure.\ref{Fig2}(b) illustrates the typical Bloch vectors $\langle \sigma_{x,y,z} (\mathbf{q}) \rangle$ in the FBZ for different quench time $\tau$ and qusiadiabatic process.
For the quasiadiabatic process, the momentum distributions of the measured Bloch vectors are consistent with those of the lowest band, indicating that none of the atoms are excited to the higher bands.
For the quench process, a region with $\langle \sigma_z(\mathbf{q}) \rangle>0$ emerges in the vicinity of the quasimomentum band closure point $M$ as a portion of the atoms are not able to flip from $\left| \uparrow \right \rangle$ to $\left| \downarrow \right \rangle$ after quench, which demonstrates adiabaticity fails.
Within this region, partial atoms are excited to the higher bands, representing the generation of the excitations.
The region contracts as the quench time increases, which means the excitations decrease.
Besides, the evolution of the Bloch vectors $\langle \sigma_{x} (\mathbf{q}) \rangle$ and $\langle \sigma_{y} (\mathbf{q}) \rangle$ forms a spiral-like pattern \cite{supM}, in which the phase of the spin wave function is determined. Meanwhile, $\langle \sigma_{z} (\mathbf{q}) \rangle$ determines the amplitude of the spin wave function.

Then the spin wave functions $|\Psi(\mathbf{q},\tau)\rangle$ are extracted from the measured Bloch vectors $\langle \bm{\sigma}(\mathbf{q},\tau) \rangle$ in Fig.\ref{Fig2} \cite{PhysRevResearch.5.L032016}: $|\Psi(\mathbf{q},\tau) \rangle=u_{\uparrow}(\mathbf{q}) \left| \uparrow \right \rangle+u_{\downarrow}(\mathbf{q}) e^{i\phi(\mathbf{q})} \left| \downarrow \right \rangle$, where $u_{\uparrow}(\mathbf{q})=\sqrt{(1+\langle\sigma_z(\mathbf{q})\rangle)/2}$ [$u_{\downarrow}(\mathbf{q})=\sqrt{(1-\langle\sigma_z(\mathbf{q})\rangle)/2}$] is the amplitude of $\left| \uparrow \right \rangle$ ($\left| \downarrow \right \rangle$) and the relative phase between $\left| \uparrow \right \rangle$ and $\left| \downarrow \right \rangle$ is $\phi(\mathbf{q})=\arg(\langle\sigma_x(\mathbf{q})\rangle+i\langle\sigma_y(\mathbf{q})\rangle)\in [-\pi,\pi)$.
The amplitudes $u_{\downarrow}(\mathbf{q})$ and the relative phases $\phi(\mathbf{q})$ for different quench time $\tau$ and qusi-adiabatic process are demonstrated in Fig.\ref{Fig3}, respectively.
Similar to the distributions of the Bloch vectors, the amplitude $u_{\downarrow}$ nearby $M$ point grows from 0 to 1 as $\tau$ increases, which means fewer atoms are excited to the higher band.
Note that the relative phase $\phi(\mathbf{q})$ is a necessary parameter for extracting excitation density, however, the excitation density is somewhat insensitive to the $\phi(\mathbf{q})$ since excitation density is dominated by the amplitudes of the spin wave function ($u_{\uparrow,\downarrow}(\mathbf{q})$).
\begin{figure}
  \includegraphics[width=0.9\linewidth]{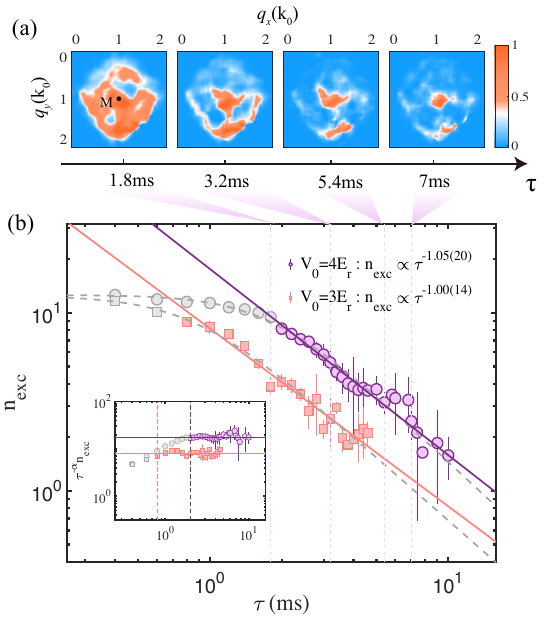}
  \caption{ Excitation density versus quench time.
  (a) The distribution of excitation density in quasimomentum space across the first Brillouin zone for different quench time $\tau$.
  Parameters: $(V_{\text{0}},\Omega_{\text{0}},\delta_i,\delta_f)=(4.0,1.0,-2.5,-0.2)E_{\rm{r}}$
  (b) The total excitation density is plotted as a function of quench time $\tau$ on log-log axes for $V_{\text{0}}=4E_{\rm{r}}$ and $V_{\text{0}}=3E_{\rm{r}}$.
  The gray circular and square symbols represent saturated regions, while the gray dashed line illustrates the fitting of the saturation model. 
  The solid purple and orange lines represent the fitting of the power law for $V_{\text{0}}=4E_{\rm{r}}$ and $V_{\text{0}}=3E_{\rm{r}}$ respectively.
  The error bars are the standard error of the mean.
  When the error bars are not visible, they are smaller than the marker size.
  The inset shows the rescaled $\tau^{-\alpha }n_{\rm exc}$, the purple and orange dashed line indicates the onset of KZ regions for $V_{\text{0}}=4E_{\rm{r}}$ and $V_{\text{0}}=3E_{\rm{r}}$ respectively.
  }\label{Fig4}
\end{figure}

With the reconstructed spin wave functions, we can obtain the power-law scaling between the total excitation density $n_{\rm exc}$ and the quench time $\tau$.
To this end, the momentum distributions of the excitation density $\rho_e(\mathbf{q},\tau)$ in the FBZ are obtained by $\rho_e(\mathbf{q},\tau)=1-|\langle \Psi(\mathbf{q},\tau) | \Psi_{\rm adi}(\mathbf{q}) \rangle|^2$, which are directly exhibited in Fig.\ref{Fig4}(a). 
Under a quench process, the atoms populated in the vicinity of the $M$ point evolve nonadiabatically and hence the region of the excitation is mainly concentrated around the $M$ point. 
Besides, the size of excitation region goes down as $\tau$ increases, which is consistent with the demonstrations for all components of the Bloch vectors and the spin wave functions.

Finally, by integrating the excitation density over the entire FBZ [see Eq.(\ref{eq2})], the total excitation density $n_{\rm exc}(\tau)$ is obtained.
We plot $n_{\rm exc}(\tau)$ as a function of $\tau$ with log-log scale for $V_0=4E_{\rm{r}}$ and $V_0=3E_{\rm{r}}$ in Fig.\ref{Fig4}(b), respectively.
The total excitation density is negatively dependent on the quench time.
As the quench time $\tau$ is increased to a certain extent, the total excitation density $n_{\rm exc}$ versus quench time $\tau$ begins to follow
a power-law scaling. 
Whereas, when the quench time decreases, the total excitation density $n_{\rm exc}$ saturates, which is also observed in other experiments exploring KZ mechanism \cite{PhysRevA.94.023628,Pyka2013}.
We distinguish the saturated region and the KZ region by fitting the experimental data using an empirical formula.
The empirical formula is $n_{\rm exc}=n_{\rm sat}[1-(\tau/ \tau_{\rm sat})^{2\beta}]^{1/2}$ \cite{PhysRevA.94.023628} with the saturation excitation density $n_{\rm sat}$, the saturation time $\tau_{\rm sat}$ and the critical exponent $\beta$.
The KZ region is determined by setting $\tau \geq 1.1\tau_{\rm sat}$ and fitted with the power function $n_{\rm exc}=C \tau^{\alpha}$, from which the critical exponent $\alpha$ is extracted.
We obtain $\alpha= -1.05\pm0.20$ ($-1.00\pm0.14$) for $V_{\text{0}}=4E_{\rm{r}}~(3E_{\rm{r}})$, which indicates that the extracted critical exponent is robust in the 2D QAH model.
For a linear quench, the band gap $\Delta$ varies linearly with detuning $\delta$.
Meanwhile, $\Delta \sim |\delta-\delta_c|^{z\nu}$ \cite{Sachdev_2011}, thus $z\nu=1$ is obtained.
Besides, for 2D system, total excitation density $n_{\text{exc}} \sim \tau^{-2\nu/(1+z\nu)}$  \cite{RN346,RN347}, hence one infers $\nu=-\alpha \approx 1$ and $z \approx 1$.
The current measured power-law scaling is similar to the mean-field KZ mechanism, providing evidence for KZ behavior in the topological transition.

Furthermore, the observed power-law scaling can be understood within the Landau-Zener (LZ) framework, where the dynamics near the critical point can decompose into a series of independent LZ transitions at each quasimomentum. The LZ probability density with quasimomentum $\mathbf{q}$ near the gap-closing point is $\rho_e(\mathbf{q},\tau)=\exp{[-(4\pi |\mathbf{q}|^2\tau/|\delta_f-\delta_i|)]}$ \cite{supM}. Integrating the excitation density over the entire FBZ, the total excitation density $n_{\rm exc} \sim\int_{\rm FBZ} \rho_e(\mathbf{q},\tau)\rm{d}\mathbf{q} \sim \tau^{-1}$, since the 2D QAH model exhibits a linear dispersion near the gap-closing point \cite{supM}. Notably, the observed power-law scaling is likely associated with the band dispersion and topological transitions \cite{supM}, which dictates the change in Chern number at the transition \cite{PhysRevB.95.075116,Chen_2019}. Typically, a linear dispersion leads to a change of $\Delta C = \pm1$ with a critical exponent $\nu = 1$ \cite{PhysRevLett.125.216601,PhysRevB.106.134203}, while quadratic dispersion tends to correspond to a change of $\Delta C = \pm2$ with $\nu = 1/2$ \cite{PhysRevB.110.165130}. These observations raise the question of whether this behavior is universal, warranting further investigation.

In summary, the measurements of the power-law scalings for the total excitation density exhibit the KZ behavior across the topological transition. Our work opens a pathway for exploring dynamic topological transitions within the KZ framework, with broad applicability to other platforms. For example, one-dimensional topological fermionic systems realized with ultracold atoms \cite{doi:10.1126/sciadv.aao4748} using our tomography technique, as well as topological dimer chains with wave-function tomography implemented in the frequency modes of an optical fiber loop platform \cite{PhysRevLett.132.183802}, provide promising avenues to verify and generalize our findings on power-law scaling behavior across topological transitions. Beyond quantifying excitation density, our experimental approach lays the groundwork for future measurements of Berry curvature in the Raman lattice. This capability enables the definition of a correlation length \cite{PhysRevLett.125.216601,PhysRevB.106.134203,Chen_2019,PhysRevB.110.165130}, thereby bridging the gap toward a full analogy with conventional symmetry-breaking transitions characterized by local order parameters. Adopting the Berry-curvature perspective grants access to dynamical features beyond excitation density alone. For example, this may enable testing the freeze-out picture of the Kibble-Zurek mechanism in topological systems  \cite{PhysRevLett.125.216601,PhysRevB.110.165130}, as well as exploring the existence of universality classes \cite{RevModPhys.39.395,RevModPhys.49.435} and coarsening dynamics (phase ordering) \cite{doi:10.1080/00018739400101505,PhysRevLett.113.095702} in topological transitions.

\begin{acknowledgments}
We thank Xiongjun Liu, Wei Zheng, Chaohong Lee, Hui Zhai and Xiaotian Nie for fruitful discussions. This work was supported by the Innovation Program for Quantum Science
and Technology (Grant No. 2021ZD0302001), the National Natural Science Foundation of China (Grant No. 12025406, 12374248 and 12104445), Anhui Initiative in Quantum Information Technologies (Grant No. AHY120000), Shanghai Municipal Science and Technology Major Project (Grant No. 2019SHZDZX01), and the Strategic Priority Research Program of Chinese Academy of Science (Grant No. XDB28000000). J.Z. acknowledges support from the CAS Talent Introduction Program (Category B) (Grant No. KJ9990007012) and the Fundamental Research Funds for the Central Universities (Grant No. WK9990000122).
\end{acknowledgments}

%

\end{document}